\documentclass[a4paper]{jpconf}
\usepackage{graphicx}
\usepackage{caption}
\usepackage{xcolor}
\usepackage{feynmp-auto}
\usepackage{subfig}
\usepackage{hyperref}
\usepackage{amsmath}
\usepackage{cleveref}
\usepackage{amssymb}
\usepackage{dutchcal} % \mathcal: also small letters. \mathbcal = bold ones.

\begin{document}
\title{Searches for additional scalars at future $e^{+}e^{-}$ colliders}

\author{Karabo Mosala$^{1}$, Thuso Mathaha$^{1}$, Mukesh Kumar$^{1}$, Bruce Mellado$^{1,2}$}

\address{$^{1}$ School of Physics and Institute for Collider Particle Physics, University of the Witwatersrand, 1 Jan Smuts Ave, Braamfontein, Johannesburg, 2000}

\address{$^{2}$ iThemba LABS, National Research Foundation, PO Box 722, Somerset West 7129, South Africa}

\ead{Karabo.mosala@cern.ch}

\begin{abstract}
We present growing excesses consistent with a $95$~GeV scalar. We provide a comprehensive analysis of the Two Higgs Doublet Model and an additional singlet (\texttt{2HDM+S}) at future $e^{+} e^{-}$ collider. In particular, we provide a precise mass reconstruction measurement for the scalar, $m_S$, using the recoil mass method through $e^{+} e^{-} \rightarrow ZS$ where $Z \rightarrow \mu^{+} \mu^{-}$ and $S \rightarrow b  \Bar{b}$ at $\sqrt{s} = 250$~GeV and $\sqrt{s} = 200$~GeV. Furthermore, we employ Deep Neural Network to analyze the properties and behaviour of the scalar particle with a mass most importantly to provide enhanced resolution for the separation between beyond the Standard Model (SM) signal and SM background in the region $95-96$~GeV in the $S \rightarrow b \Bar{b}$ for $\mu^{+} \mu^{-}$ channel. A $95$~GeV scalar can be observed with $5 \sigma$ significance at $15(10)$ fb$^{-1}$ integrated luminosity for $\sqrt{s} = 250(200)$~GeV. This strengthens the discovery of the potential of the future $e^{+} e^{-}$ collider.
\end{abstract}

\section{Introduction}
\label{sec:intro}
 The proposal for the development of $e^+ e^-$ collider provides a window of opportunity for the field of particle physics. This colliders, equipped with advanced detectors, will provide valuable data for these studies. In previous works \cite{vonBuddenbrock:2015ema, Kumar:2016vut}, the introduction of scalars $H$ and $S$ in an effective model was proposed to explain various features observed in the collider data. These features includes distortions in the Higgs boson ($h$) elevated associated jet activity, transverse momentum spectrum, and elevated lapton rates in association with $b$-tagged jets. Additionally, the search for double Higgs boson and weak boson production has also been explored. 

Extensions of the standard model (SM) include the \texttt{2HDM}, which introduce an additional Higgs doublet. This leads to the presence of two CP-even ($h, H$) scalar bosons, one CP-odd ($A$) scalar boson, and charged scalar bosons ($H^{\pm}$). While \texttt{2HDM} have been extensively studied in the literature, it has been pointed out that a \texttt{2HDM} alone cannot explain all the observed features in the data. Therefore, a scalar singlet $S$ has been introduced in conjunction with the \texttt{2HDM}, resulting in the \texttt{2HDM+S} model. This model also provides a framework to explore scenarios involving dark matter \cite{Bhattacharya:2023qfs}.

In this study, we build upon the phenomenology described in previous works and investigate the \texttt{2HDM+S} parameter space that can accommodate the observed anormalies in the data \cite{vonBuddenbrock:2015ema}. Specifically, we focus on the production of the singlet scalar ($S$) through the $e^{+} e^{-}$ production mode, with $S \rightarrow b \Bar{b}$ channel.

The CMS experiment at the Large Hadron Collider (LHC) reported a local excess of 2.9$\sigma$ at 95.4 GeV in the di-photon invariant mass spectrum using Run 2 data~\cite{ATLAS:2013dos,Bahl:2022igd}. The ATLAS experiment has recently reported a much smaller excess of 1.7$\sigma$ at the same mass with the same data set~\cite{Butterworth:2010ym}. A local excess of 2.8$\sigma$ at 95.3 GeV was reported by the CMS experiment in the $\tau^{+} \tau^{-}$ final states~\cite{ENGELN2019256}.

\section{Model and Framework}
\label{sec:model}
A Two Higgs doublet model (\texttt{2HDM}) and the addition of a real singlet $\Phi_S$ forms the baseline for the \texttt{2HDM+S} formalism. Within \texttt{2HDM+S} framework, the potential is given by:
\begin{equation}
\label{eq:1}
\begin{aligned}
V(\Phi_{1},\Phi_{2},\Phi_{S}) &= m_{11}^2|\Phi_1|^{2} + m_{22}^2|\Phi_2|^{2} - m_{12}^2(\Phi_1^{\dagger} \Phi_2 + h.c.) + \frac{\lambda_1}{2} (\Phi_1^{\dagger}\Phi_1) + \frac{\lambda_2}{2} (\Phi_{2}^{\dagger}\Phi_2) \\
&\quad + \lambda_{3}(\Phi_1^{\dagger}\Phi_1)(\Phi_2^{\dagger}\Phi_2) + \lambda_{4}(\Phi_1^{\dagger}\Phi_2)(\Phi_2^{\dagger}\Phi_1) + \frac{\lambda_5}{2} [(\Phi_1^{\dagger}\Phi_2)^{2} + h.c.] \\
&\quad + \frac{1}{2} m_{S}^{2} \Phi_{S}^{2} + \frac{\lambda_6}{8} \Phi_{S}^{4} + \frac{\lambda_7}{2} (\Phi_1^{\dagger}\Phi_1) \Phi_{S}^{2} + \frac{\lambda_8}{2} (\Phi_2^{\dagger}\Phi_2) \Phi_{S}^{2}.
\end{aligned}
\end{equation}
where $\Phi_1$ and $\Phi_2$ represent $SU(2)_L$ Higgs doublets. The first two lines pertain to the 2HDM potential, while the final line introduces input from the singlet field $\Phi_S$. After electroweak symmetry breaking, the two doublet fields acquire the real VEVs $v_1$ and $v_2$ and the singlet field a real VEV $v_S$ and they can be parametrised as

\begin{equation}
\label{eq:15}  
\Phi_{1} = 
    \begin{pmatrix}
\phi^{+}_{1} \\
\frac{1}{\sqrt{2}} \left(v_1 + \rho_1 + i\eta_1 \right)
\end{pmatrix}, \Phi_{2} = 
    \begin{pmatrix}
\phi^{+}_{2} \\
\frac{1}{\sqrt{2}} \left(v_2 + \rho_2 + i\eta_2 \right)
\end{pmatrix} , \Phi_{S} = v_{S} + \rho_{S},
\end{equation}
given in terms of a real neutral CP-add and CP-even fields $\rho_{I} (I = 1, 2, S)$ and $\eta_{i}$ and complex charged fields $\phi^{+}_{i}, (i = 1, 2)$, respectively. Within the \texttt{2HDM+S} model, we consider the following set of input parameters
\begin{equation}
    \begin{array}{cc}
        m_{H_{1,2,3}},\; m_{1,2}^{2}, \; m_{H^{\pm}},\; m_{A},\; \alpha_{1},\; \alpha_{2},\;  \alpha_{3},\; t_{\beta},\; v_{S},\; v.
        \label{Scan96}
    \end{array}
\end{equation}
We refer the reader to Ref.\cite{vonBuddenbrock:2018xar,Muhlleitner:2016mzt} for further details.

\section{Analysis, Observable and results}
\label{sec:Results}
\subsection{Discovery potential of CEPC}
\label{subsec:recoil}
To assess the capability of the future Circular Electron-Positron Collider(CEPC) in terms of precision measurements related to the Higgs boson, an extensive set of benchmark studies in Higgs physics is underway. These studies encompass detector simulations that incorporate various factors, such as pile-up resulting from background processes like $\gamma \gamma \rightarrow$ hadron interactions, as well as the pertinent background processes from $e^{+} e^{-}$ collisions. The simulations are designed to consider the unique CEPC beam spectrum and account for effects like initial state radiation that are relevant to the analysis.

The analysis of events originating from the Higgsstrahlung process $e^{+} e^{-} \rightarrow Z S$ serves as a significant method for precision measurements within the context of Higgs physics. This process, dominant at center-of-mass energies below $\sqrt{s} = 250$ GeV, provides a platform for two crucial measurements. Firstly, it enables the precise determination of the coupling between the Higgs boson and the $Z$ boson. Secondly, it facilitates the measurement of the interactions between the Higgs boson and its final decay products.

Through this approach, an unbiased assessment of the coupling $g_{HZZ}$ is carried out, independent of specific models. This involves reconstructing the recoil mass of the $Z$ boson, contributing to the precision of the measurement~\cite{Chen:2016zpw}
\begin{align}
\label{eq:3.2}
   {\rm M_{recoil}} = \sqrt{s + M_{\mu^{+} \mu^{-}}^2 - 2(E_{\mu^{+}} + {E_{\mu^{-}}})\sqrt{s}},
\end{align}
with no reconstruction of the Higgs necessary. $\sqrt{s}$ is the centre of mass energy, $M_{\mu^{+} \mu^{-}}^2$,\; $E_{\mu^{+}} + {E_{\mu^{-}}}$ is the invariant mass and energy of muons, respectively.

To generate SM backgrounds and BSM signals, we specify $\sqrt{s}$, beam polarization ($e^{+}$ and $e^{-}$ beams are 80\% left polarized and 80\% right polarized, respectively), and $\mathcal{L}$, as well as considering $95$~GeV scalar-production modes and their corresponding cross sections for $E_{e^{+}} = 100/125$~GeV and $E_{e^{-}} = 100/125$~GeV. To optimise the signal events over the leading backgrounds minimal cuts on leading and sub-leading jets, $j\;j$ and leading and sub-leading muons, $\mu^{+} \mu^{-}$ and leading and sub-leading $b$-tagged jets, $b \Bar{b}$ are applied for all processes in this study with appropriate additional cuts; $E_{j,b,\mu}  > 5$~GeV, $\mathrm{M}^{\mathrm{Recoil}}_{\mu_{+} \mu_{-}} < 120$~GeV, $\mathrm{M}_{b \Bar{b}} < 100$~GeV.

\begin{figure}[t]
\begin{center}
\includegraphics[width=0.40\textwidth]{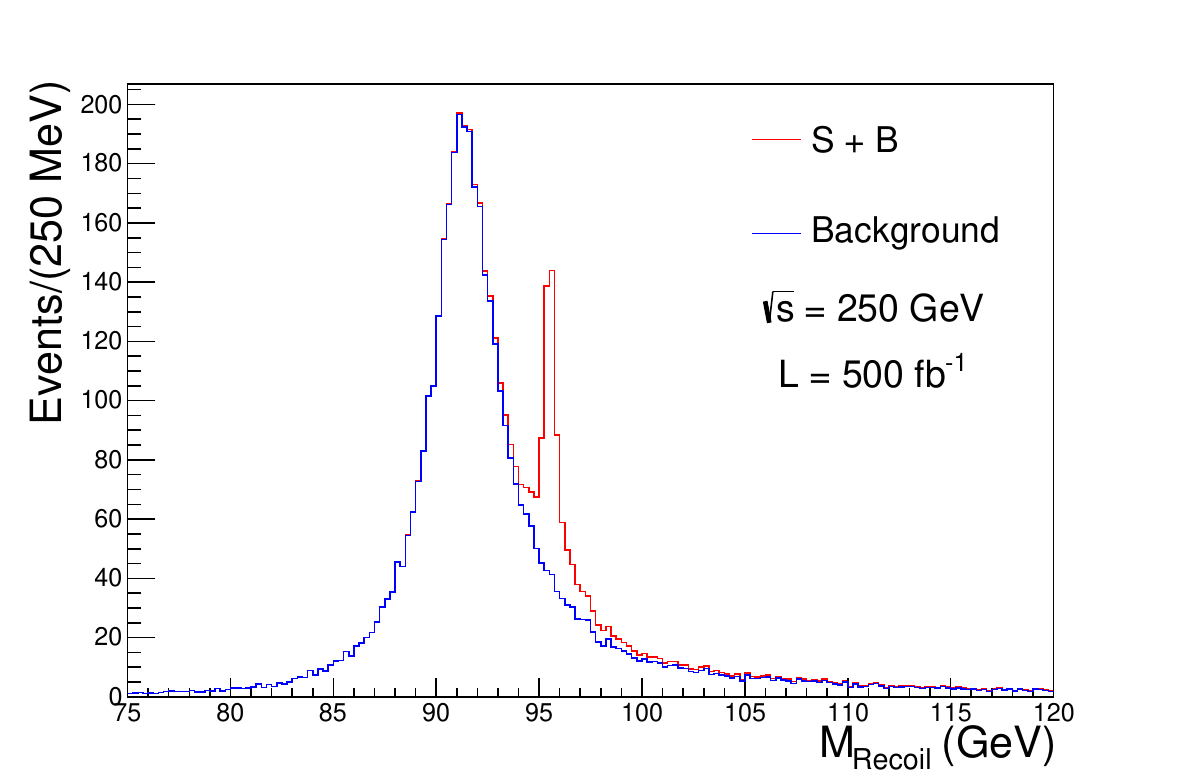}
\includegraphics[width=0.40\textwidth]{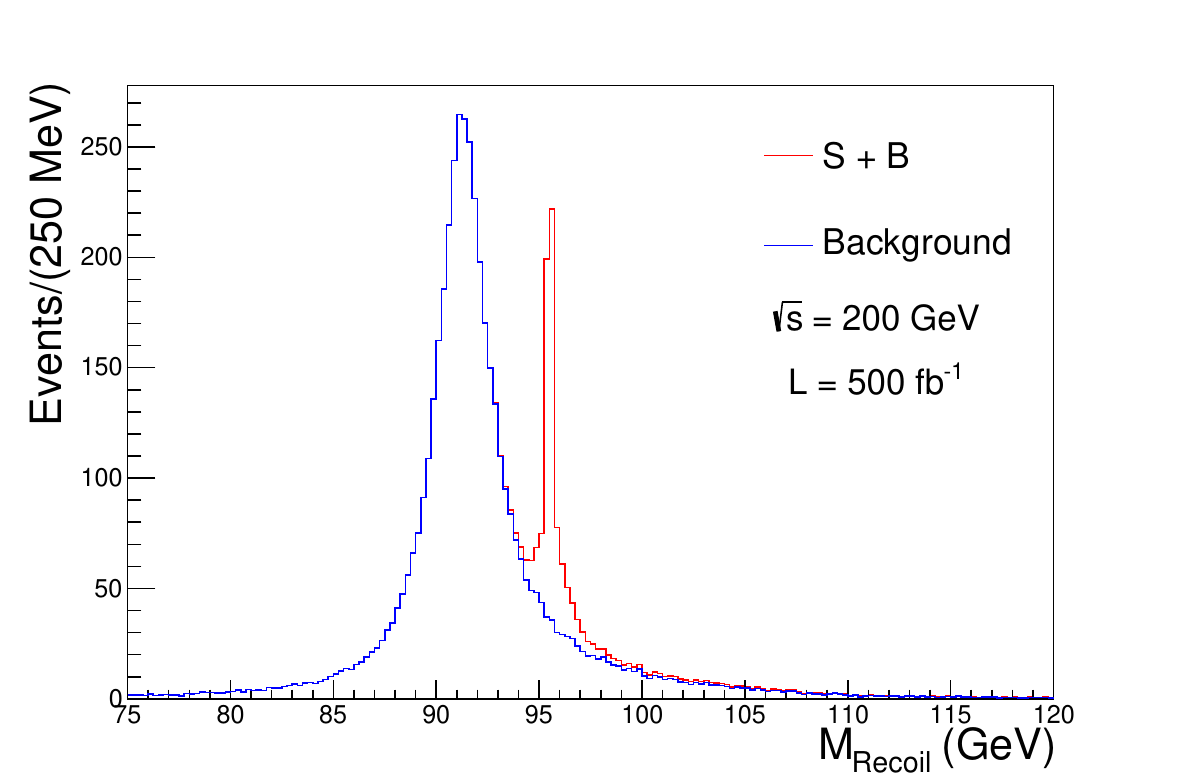}
\caption{Recoil mass distribution for
$e^{+} e^{-} \rightarrow S Z \rightarrow S \mu^{+} \mu^{-}$ events with $m_{S} = 95$~GeV at integrated luminosity $\mathcal{L}= 500$ fb$^{-1}$ for $\sqrt{S}=250$~GeV (left) and $\sqrt{S}=200$~GeV (right). Red is S$+$B where S takes into consideration 10\% SM Higgs signal, B is the SM background and Blue is the SM background.}
\label{fig:6}
\end{center}
\end{figure}

%\newpage
\subsection{Signal and Background Discrimination}
\label{subsec:DNN}
\begin{figure}[t]
\centering
\includegraphics[width=0.32\textwidth]{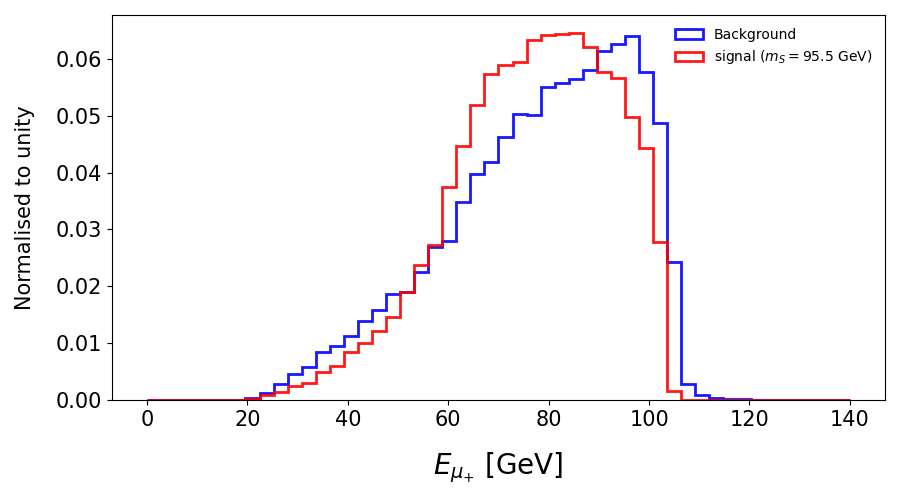}
\includegraphics[width=0.32\textwidth]{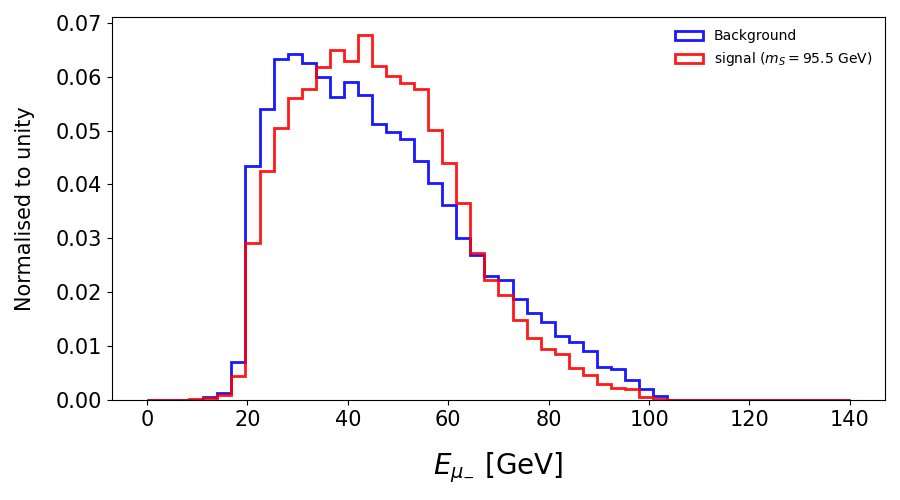}
\includegraphics[width=0.32\textwidth]{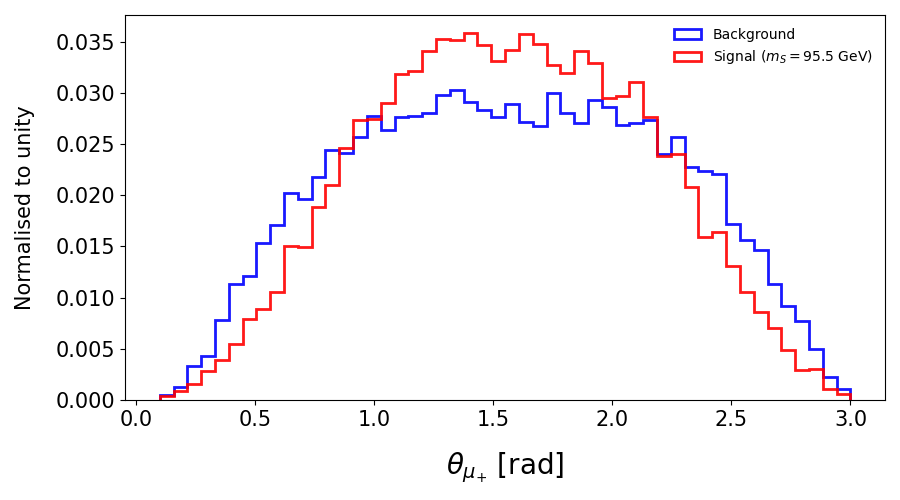} \\
\includegraphics[width=0.32\textwidth]{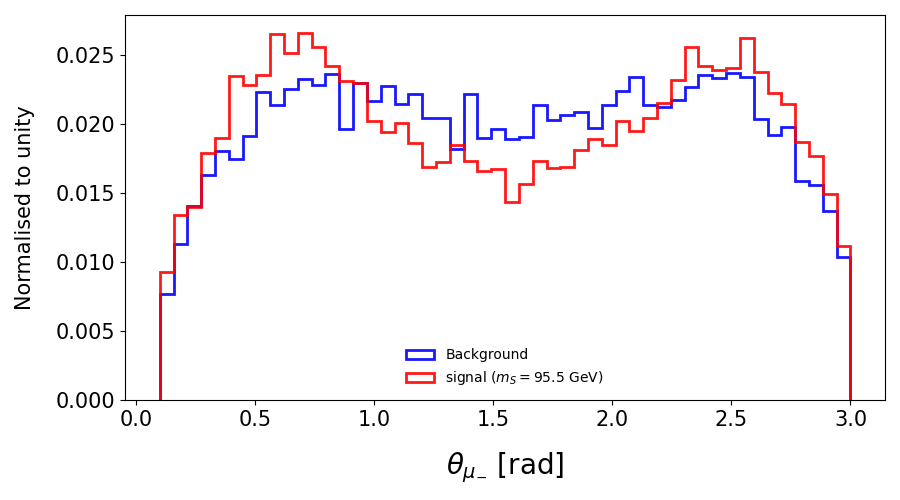}
\includegraphics[width=0.32\textwidth]{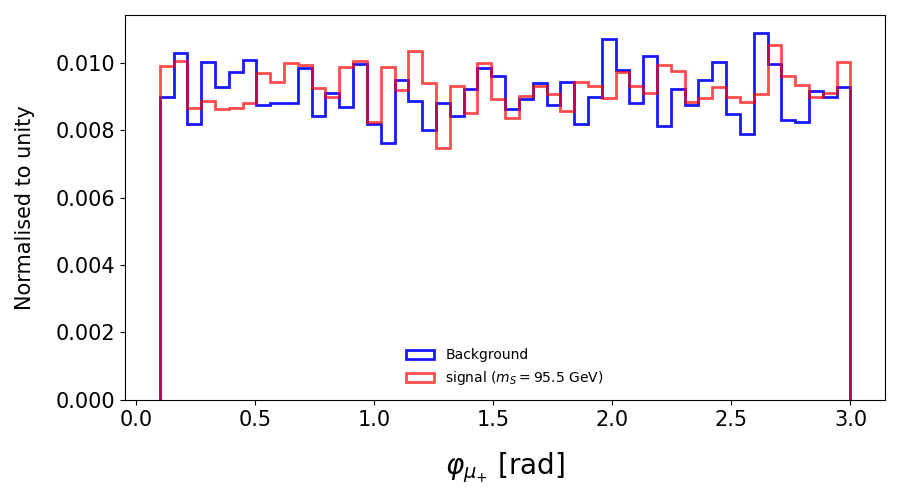}
\includegraphics[width=0.30\textwidth]{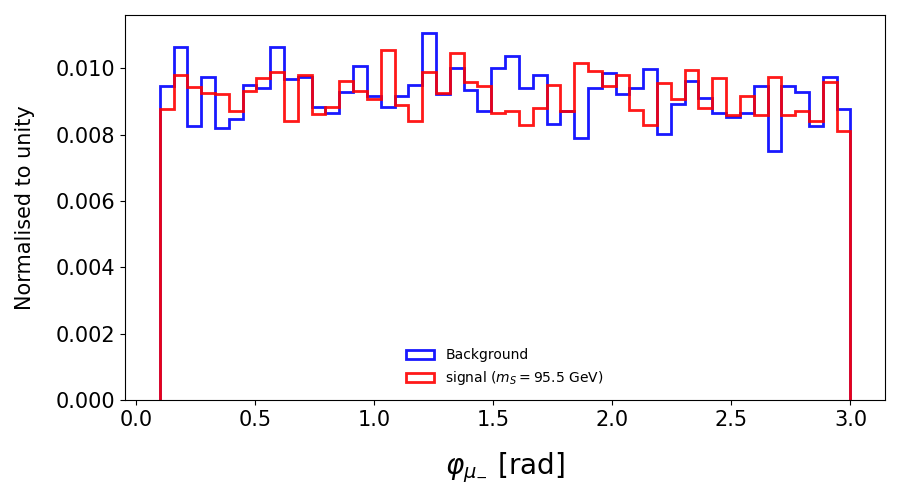} \\
\includegraphics[width=0.30\textwidth]{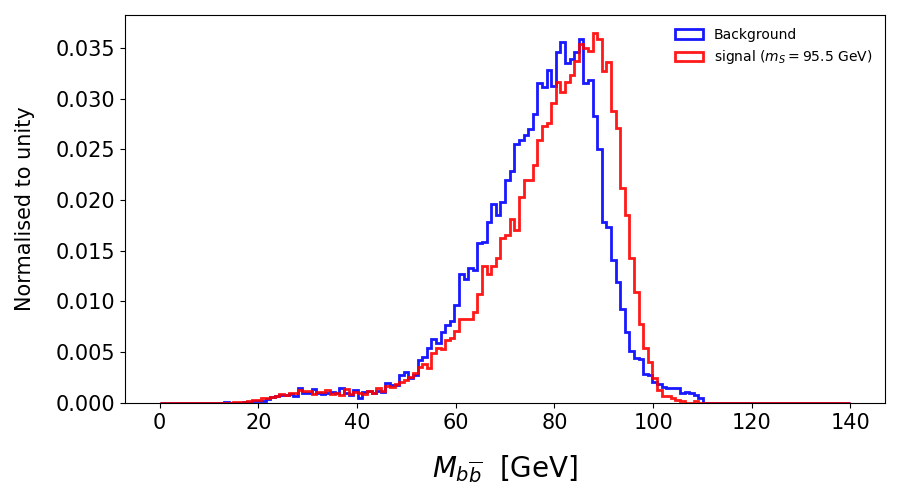}
\includegraphics[width=0.30\textwidth]{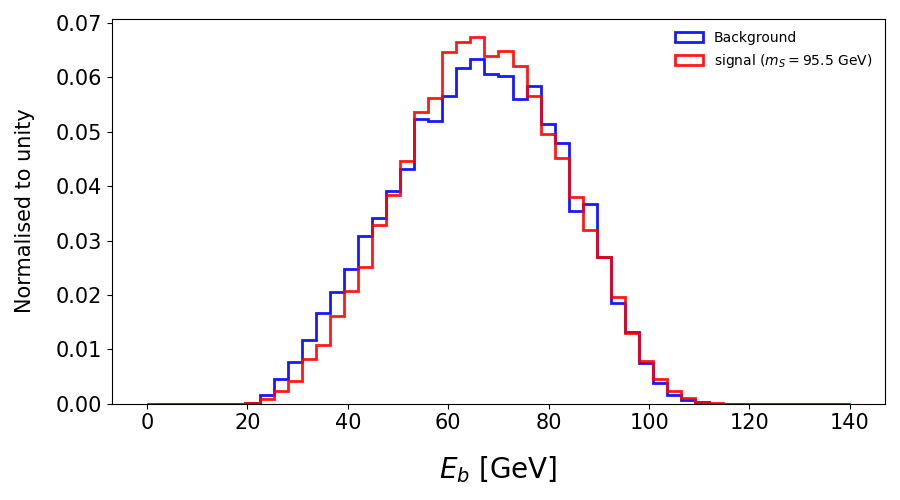}
\includegraphics[width=0.30\textwidth]{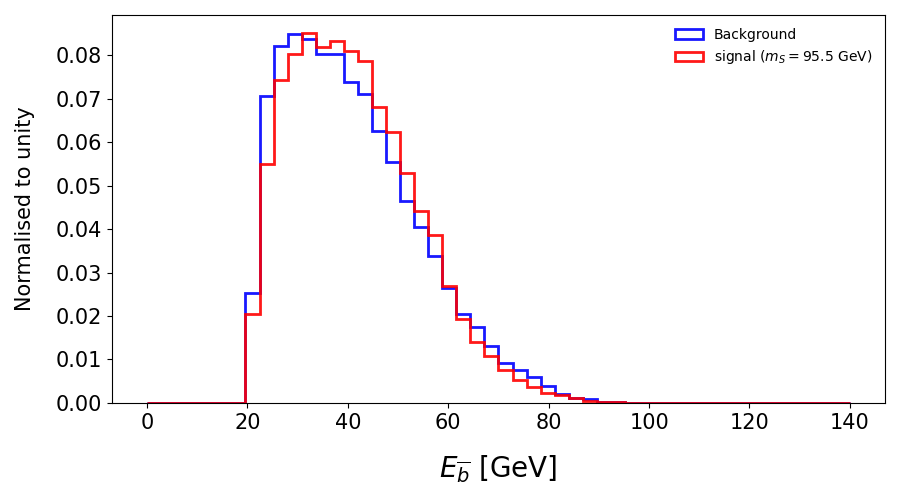}
\includegraphics[width=0.30\textwidth]{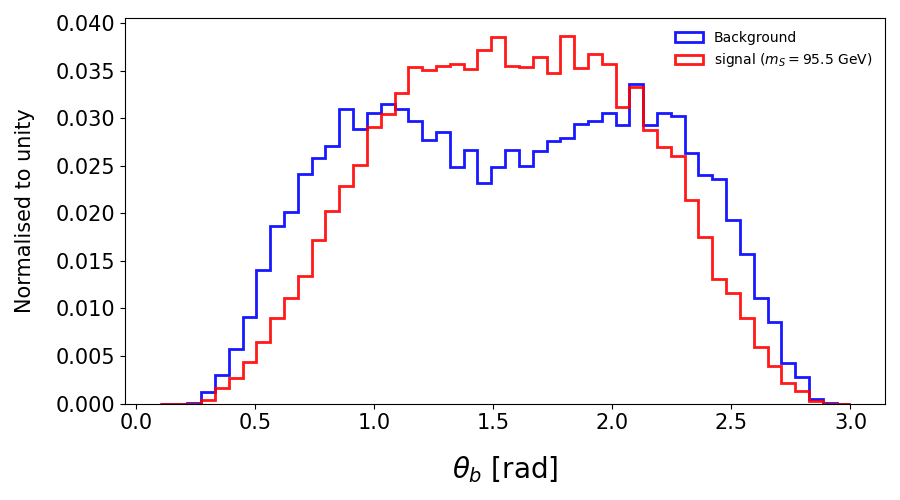}
\includegraphics[width=0.30\textwidth]{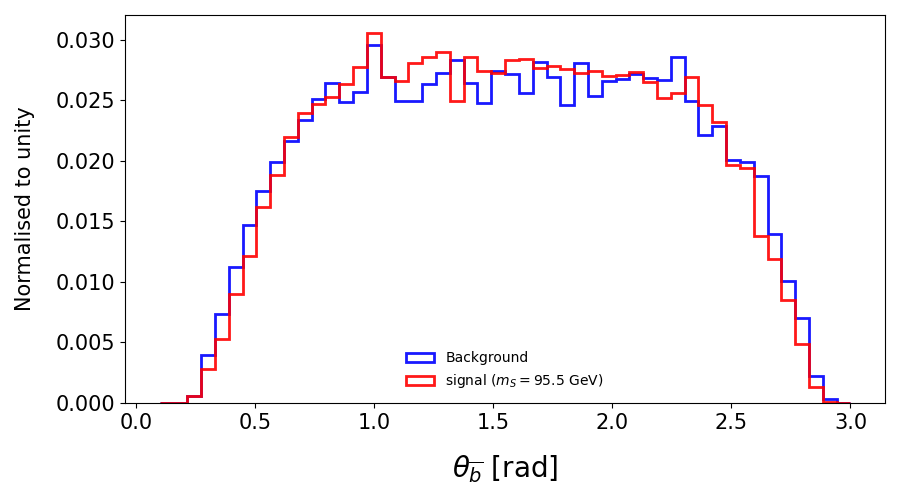}
\includegraphics[width=0.30\textwidth]{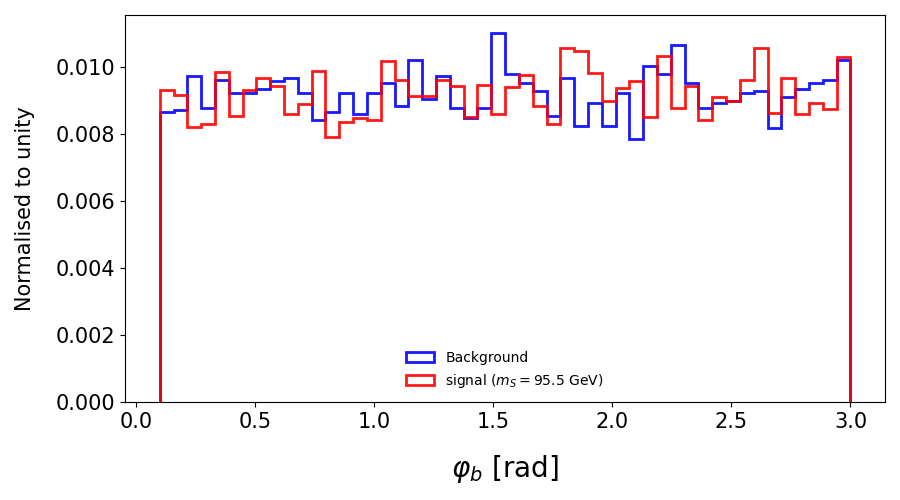}
\includegraphics[width=0.30\textwidth]{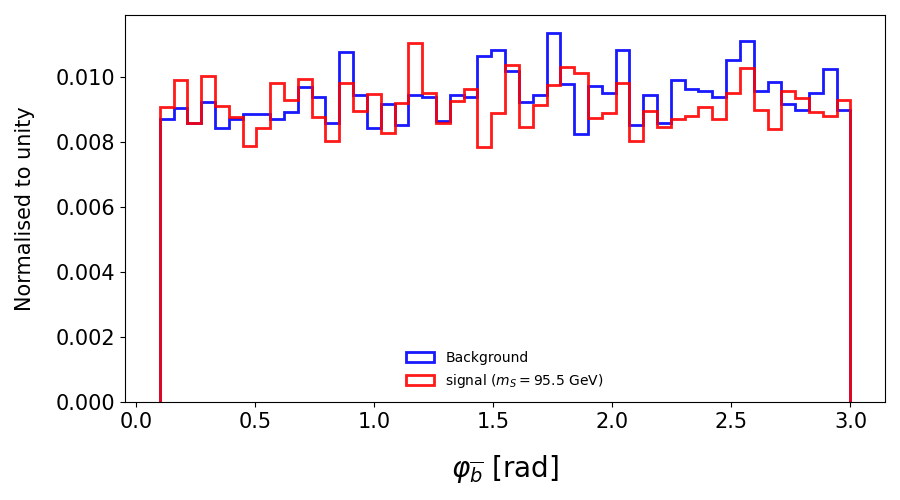}
\caption{Chosen sensitive kinematic distributions for the muon pair ($\mu_{+} = \mu_{1}$, $\mu_{-} = \mu_{2}$) and the $b$-tagged jets used as input variables for DNN normalized to unity at $\sqrt{s}= 250$~GeV.}
\label{fig:8}
\end{figure}

The event selection criteria described in Section \ref{subsec:recoil} eliminate most of the background events coming from different SM physics processes. However, the number of remaining background events after this selection may still have a substantial dilution effect. Therefore, a second selection step was developed by training a Machine Learning (ML) model to further discriminate the signal events from the background events. A SM sample for $e^{+} e^{-} \rightarrow Z S, \;S \rightarrow b\overline{b}, \;Z \rightarrow \;\mu^{+} \mu^{-}$ together with BSM sample $e^{+} e^{-} \rightarrow Z b \Bar{b},\; Z \rightarrow \mu^{+} \mu^{-}$ has been generated to be used in the training. The training was performed by using several event and physics object variables as the inputs of the model Figure~\ref{fig:8}.

\begin{figure}[t]
\centering
\includegraphics[width=0.40\textwidth]{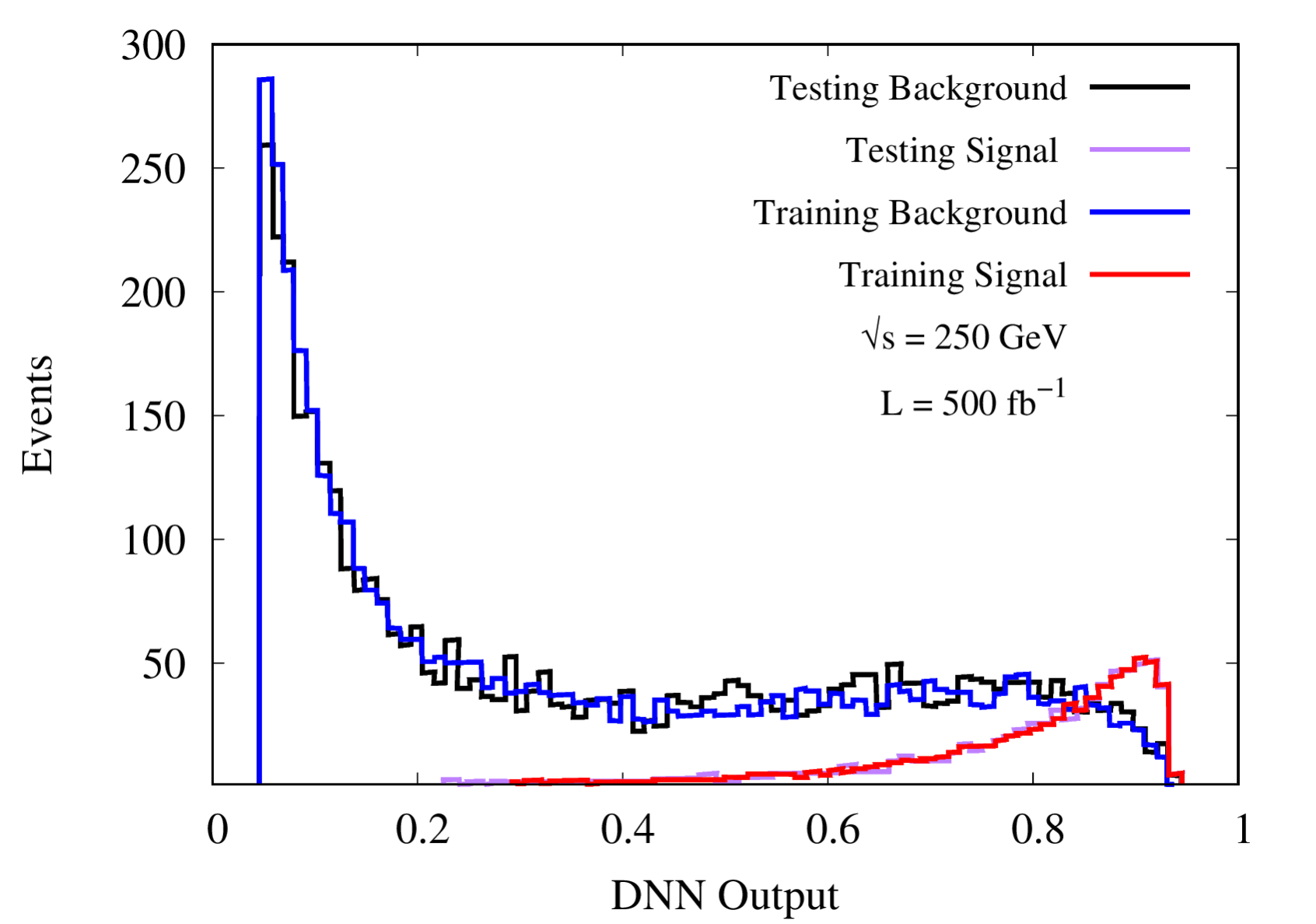}
\includegraphics[width=0.40\textwidth]{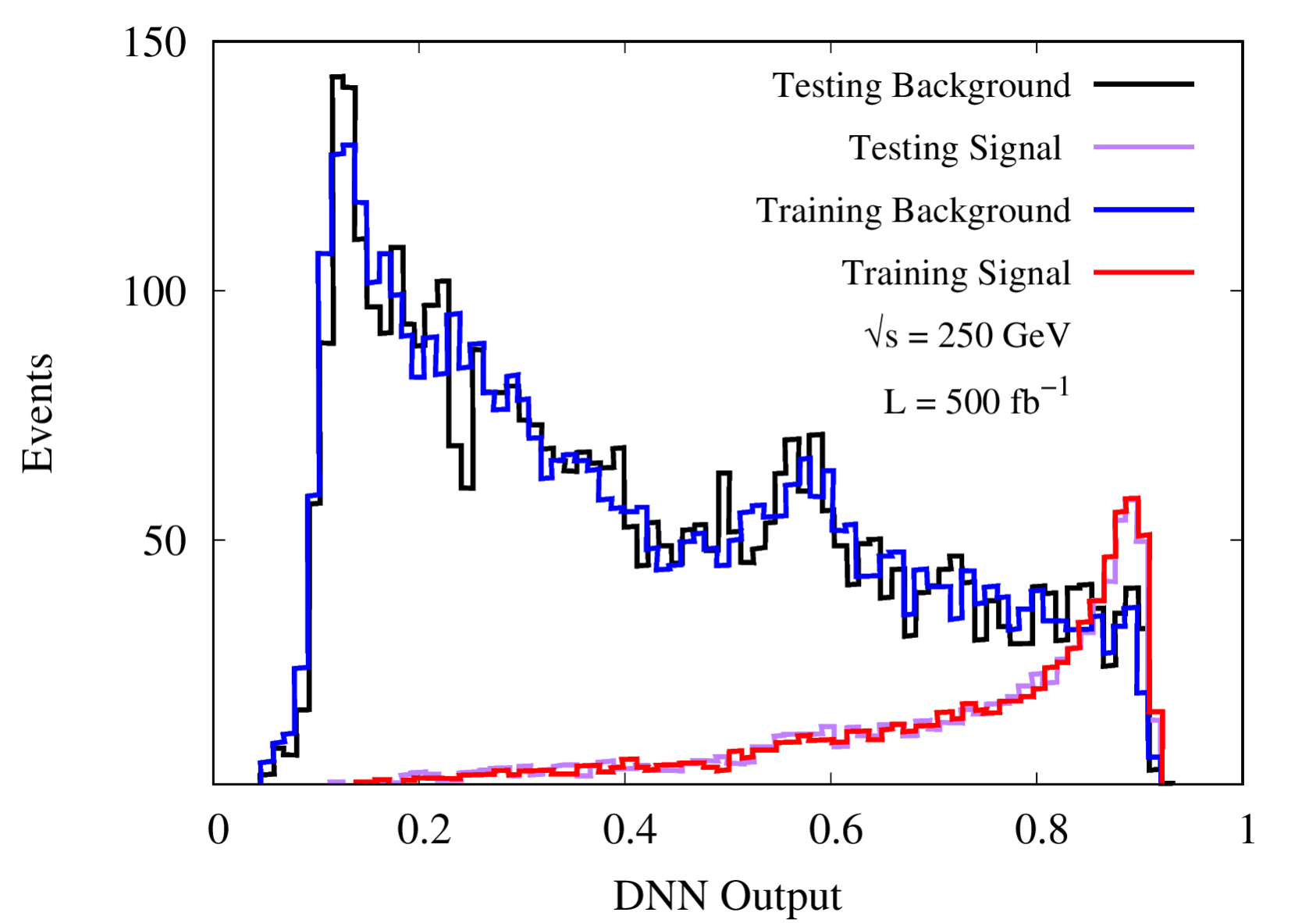}
\caption{The DNN output with training and testing when the recoil mass is not included (right) and when the recoil mass is included (left) and the respective ROC curves below for $\sqrt{s}= 250$~GeV.}
\label{fig:10}
\end{figure}

In Figure~\ref{fig:10}, the normalized deep neural network (DNN) model output distributions for both signal and background events separately on the training and validation samples are shown. A significant difference has been observed between the distributions obtained from the training and the validation samples for both signal and background events with the inclusion (left) and exclusion (right) of the recoil mass respectively.

\begin{figure}[htp]
\centering
\includegraphics[width=0.40\textwidth]{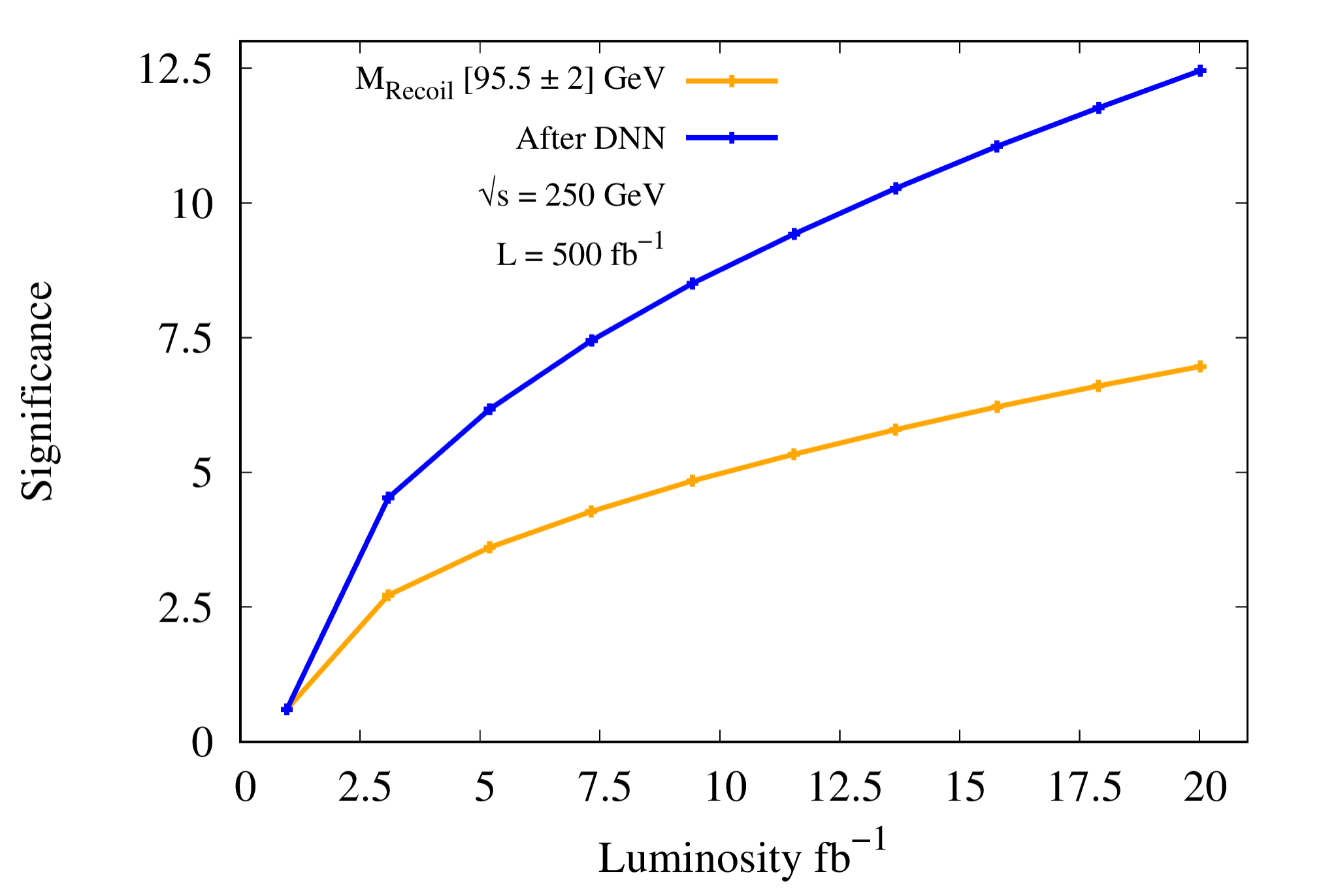}
\includegraphics[width=0.40\textwidth]{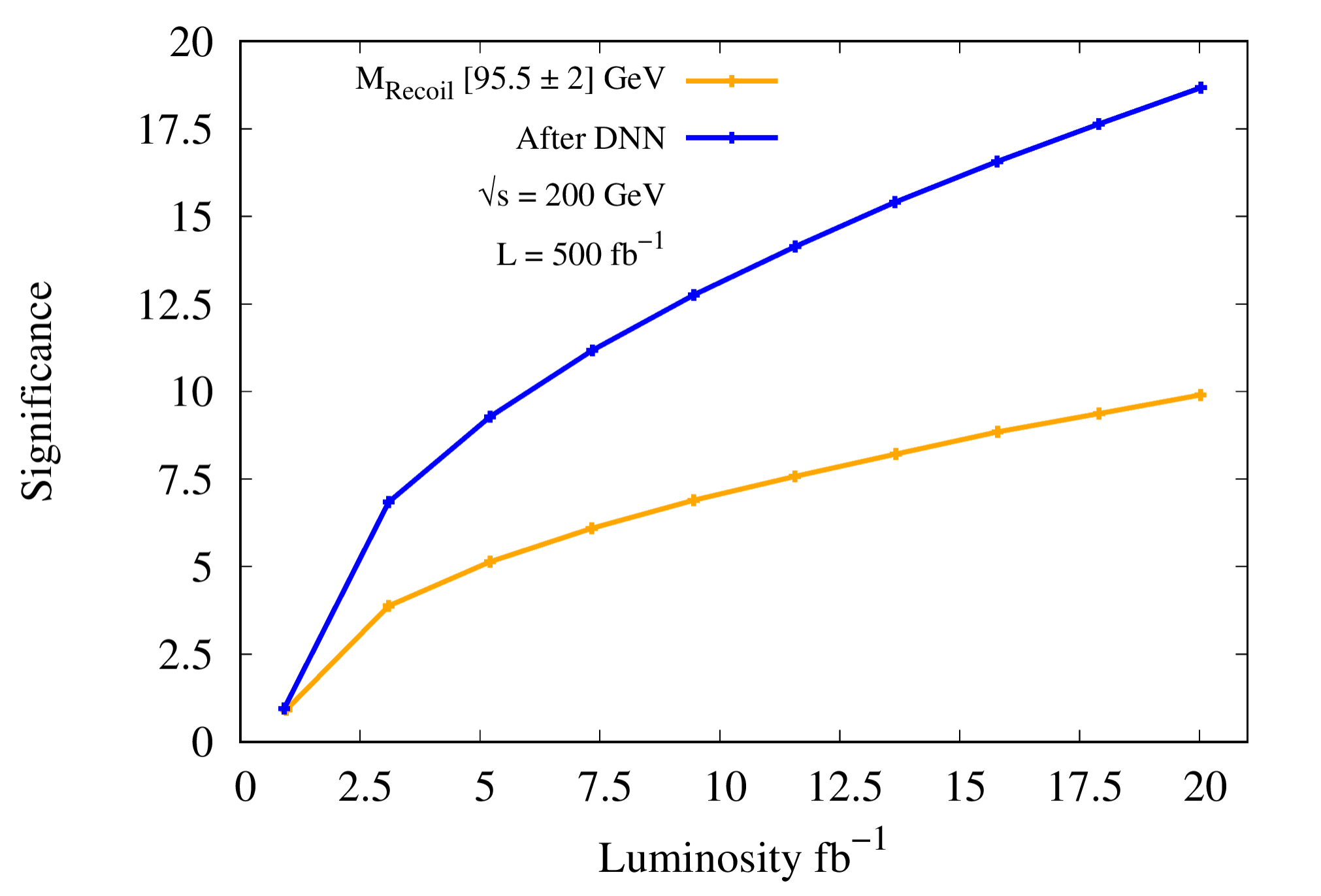}
\caption{The signal significance as a function of Luminosity ($\mathcal{L}$) when recoil mass window of $\pm 2$ GeV is considered and after DNN for (left) $\sqrt{s}=250$ and (right) $\sqrt{s}=200$ at $\mathcal{L} = 500$ fb$^{-1}$ integrated luminosity at CEPC.}
\label{fig:11}
\end{figure}
\newpage
\section{Conclusions}
\label{sec:Sum}
This proceedings investigated the potential for producing the $\approx 95$~GeV Scalar ($S$) at $\sqrt{s}=250$~GeV and $\mathcal{L} = 500$ fb$^{-1}$ integrated luminosity CEPC environment. Figure~\ref{fig:6} presents recoil mass reconstruction spectrum. DNN was crucial for a clear distinction in the region $m_S \in [90 - 96]$~GeV at $\sqrt{s}= 250$~GeV and $\sqrt{s} = 200$~GeV. Figure~\ref{fig:11} displays optimized signal significance relevant to CEPC initial operation at considering 5\% systematics. %Furthermore, we are going to provide the initial CEPC Scalar production measurement reconstruction at the DNN predicted luminosity $\mathcal{L}= 40 $ fb$^{-1}$.

%\newpage


\section*{References}
\begin{thebibliography}{9}
%\cite{vonBuddenbrock:2015ema}
\bibitem{vonBuddenbrock:2015ema}
Buddenbrock S~V, Chakrabarty N, Cornell A~S, Kar D, Kumar M, Mandal T, Mellado B, Mukhopadhyaya B and Reed R~G
%``The compatibility of LHC Run 1 data with a heavy scalar of mass around 270\textbackslash{},GeV,''
(preprint 1506.00612)
%47 citations counted in INSPIRE as of 10 Jul 2023


%\cite{Kumar:2016vut}
\bibitem{Kumar:2016vut}
Kumar M, Buddenbrock S~V, Chakrabarty N, Cornell A~S, Kar D, Mandal T, Mellado B, Mukhopadhyaya B and Reed R~G 2017
%``The impact of additional scalar bosons at the LHC,''
J. Phys. Conf. Ser. \textbf{802} no.1 012007
%doi:10.1088/1742-6596/802/1/012007
%(preprint 1603.01208)
%12 citations counted in INSPIRE as of 10 Jul 2023

%\cite{Bhattacharya:2023qfs}
\bibitem{Bhattacharya:2023qfs}
Bhattacharya S, Dey A, Lahiri J and Mukhopadhyaya B 
%``High scale validity of two Higgs doublet scenarios with a real scalar singlet dark matter,''
(preprint 2308.12473)
%0 citations counted in INSPIRE as of 11 Oct 2023

\bibitem{ATLAS:2013dos}
Georges A~\textit{et al.}
(ATLAS) 2013
Phys. Lett. B 726 88-119

\bibitem{Bahl:2022igd}
Bahl H, Biek\"otter T, Heinemeyer S, Li C, Paasch S, Weiglein G and Wittbrodt J 2023
%``HiggsTools: BSM scalar phenomenology with new versions of HiggsBounds and HiggsSignals,''
Comput. Phys. Commun. \textbf{291} 108803
%doi:10.1016/j.cpc.2023.108803
%(preprint 2210.09332)
%26 citations counted in INSPIRE as of 25 Aug 2023

%\cite{Butterworth:2010ym}
\bibitem{Butterworth:2010ym}
Butterworth J~M, Arbey A, Basso L, Belov S, Bharucha A, Braam F, Buckley A, Campanelli M, Chierici R and Djouadi A \textit{et al.} 2009
%``THE TOOLS AND MONTE CARLO WORKING GROUP Summary Report from the Les Houches 2009 Workshop on TeV Colliders,''
(preprint 1003.1643)
%199 citations counted in INSPIRE as of 25 Aug 2023

\bibitem{ENGELN2019256}
Isabell E,  Margarete M, and Jonas W  2019 Comp. Phy. Com. 234 256-262 


%\cite{vonBuddenbrock:2018xar}
\bibitem{vonBuddenbrock:2018xar}
Buddenbrock S~V, Cornell A~S, Iarilala E~D~R, Kumar M, Mellado B, Ruan X and Shrif E~M 2019
%``Constraints on a 2HDM with a singlet scalar and implications in the search for heavy bosons at the LHC,''
J. Phys. G \textbf{46} no.11 115001
%doi:10.1088/1361-6471/ab3cf6
%(preprint 1809.06344)
%30 citations counted in INSPIRE as of 25 Aug 2023


%\cite{Muhlleitner:2016mzt}
\bibitem{Muhlleitner:2016mzt}
Muhlleitner M, Sampaio M~O~P, Santos R and Wittbrodt J 2017
%``The N2HDM under Theoretical and Experimental Scrutiny,''
JHEP \textbf{03} 094
%doi:10.1007/JHEP03(2017)094
%[arXiv:1612.01309 [hep-ph]].
%107 citations counted in INSPIRE as of 26 Oct 2023

%\cite{Chen:2016zpw}
\bibitem{Chen:2016zpw}
Chen Z, Yang Y, Ruan M, Wang D, Li G, Jin S and Ban Y 2017
%``Cross Section and Higgs Mass Measurement with Higgsstrahlung at the CEPC,''
Chin. Phys. C \textbf{41} no.2 023003
%doi:10.1088/1674-1137/41/2/023003
%[arXiv:1601.05352 [hep-ex]].
%28 citations counted in INSPIRE as of 26 Oct 2023


\end{thebibliography}
\end{document}